\documentstyle[aps,twocolumn,epsfig]{revtex}
\catcode`\@=11

\newcommand{\fcaption}[1]{\vspace{1ex}
        \refstepcounter{figure}
        \setbox\@tempboxa = \hbox{\footnotesize {\bf ig.~\thefigure.} #1}
        \ifdim \wd\@tempboxa > 8cm
           {\begin{center}
        \parbox{8cm}{\footnotesize\baselineskip=8pt {\bf Fig.~\thefigure.} #1}

            \end{center}}
        \else
             {\begin{center}
             {\footnotesize {\bf Fig.~\thefigure.} #1}
              \end{center}}
        \fi}

\begin{document}

\title{Dynamical Symmetry Enlargement Versus
Spin-Charge Decoupling in the One-Dimensional SU(4) Hubbard Model}
\vspace{2cm}

\author{R. Assaraf $^1$, P. Azaria $^2$, E. Boulat $^{2,3}$,
M. Caffarel $^1$, and P. Lecheminant $^4$}  

\vspace{0.5cm}

\address{
$^1$ CNRS-Laboratoire de Chimie Th{\'e}orique, Universit{\'e} Paris 6 , 4 Place Jussieu, 75252 Paris Cedex 05, France \\
$^2$CNRS-Laboratoire de Physique Th{\'e}orique des Liquides,
Universit{\'e} Paris 6, 4 Place Jussieu, 75252 Paris Cedex 05,
France\\
$^3$ Center For Materials Theory, Serin Physics Laboratory,
Rutgers University, Piscataway, New Jersey 08854-8019, USA \\
$^4$ Laboratoire de Physique Th\'eorique et
Mod\'elisation, CNRS UMR 8089,
Universit\'e de Cergy-Pontoise, 5 mail Gay-Lussac, Neuville sur Oise,
95301 Cergy-Pontoise Cedex, France
}
\vspace{3cm}

\address{\rm (Received: )}
\address{\mbox{ }}
\address{\parbox{14cm}{\rm \mbox{ }\mbox{ }
We investigate  dynamical symmetry enlargement in the 
half-filled SU(4) Hubbard chain using non-perturbative 
renormalization group and Quantum
Monte Carlo techniques. A spectral gap is shown to open 
for arbitrary Coulombic repulsion $U$.
At weak coupling, $U \lesssim 3t$, a SO(8) symmetry
between charge and spin-orbital excitations is found to be 
dynamically enlarged at low energy. At strong
coupling, $U \gtrsim 6t$, the charge degrees of freedom 
dynamically decouple and the resulting 
effective theory in the spin-orbital sector is that of 
the SO(6) antiferromagnetic 
Heisenberg model. Both regimes exhibit spin-Peierls order. However, 
 although spin-orbital excitations are $incoherent$  
 in the SO(6) regime they are   $coherent$ in the SO(8) one.
The cross-over between these regimes is 
discussed.
}}
\address{\mbox{ }}
\address{\parbox{14cm}{\rm PACS No: 71.10.Fd, 71.10.Pm, 02.70.Ss, 71.30.+h}}
\maketitle
In strongly correlated electronic systems the presence of additional 
dynamical degrees
of freedom out of the usual spin and charge ones is expected to play
an important role in a number of complex systems.
This is the case, for example, of some $d$-electron 
systems\cite{kugel}, 
C$_{60}$-based materials\cite{arovas} and, also, various 
ladders-type compounds\cite{kugelspinorbital},
for which low-energy excitations cannot be constructed from 
a single effective orbital per site (one band).
An  important question that arises is to know 
whether or not there exist $generic$ 
features associated with multi-orbital effects. Such a question is non-trivial
since it is known that the lack of symmetry in multi-orbital 
problems (beyond the usual SU(2) spin-invariance)
is responsible for the presence of many  independent couplings 
and, therefore, a wide range of problem-dependent
physical behaviors could be expected. However, at sufficiently low energy, 
it may happen
that the effective symmetry is increased, thus
considerably simplifying the description  of the problem. 
This is of course what  happens in  a critical (gapless) model.
In the more general case  where  a spectral gap is present, 
the possibility of such a Dynamical Symmetry 
Enlargement (DSE) at low energy  is clearly  non  trivial.   
Recently, 
Lin, Balents, and Fisher\cite{lin} 
have emphasized  that DSE is likely to be a generic 
tendency of  the perturbative (one-loop)  Renormalization-Group (RG) 
flow in their study of the half-filled two-leg Hubbard ladder.
However, since in a gapped system  DSE is a $strong$ coupling 
effect  one may thus question the reliability
of perturbation theory\cite{azaria0}. Clearly, in view of the
importance that such a DSE phenomenon might have
in our understanding of complex systems, 
a non-perturbative investigation is called for and it 
is the purpose of this Letter to present such a study.

In the following, we investigate the DSE phenomenon using 
non-perturbative RG and Quantum Monte Carlo (QMC)
simulations for the simplest one-dimensional half-filled two-band Hubbard model 
where spin and orbital degrees of freedom play a $symmetrical$ role. The 
corresponding SU(4) Hubbard model reads as follows:
\begin{equation}
{\cal H} = - t\; \sum_{i,a\sigma}  \left(c_{i,a\sigma}^{\dagger}
c_{i+1,a\sigma} + {\rm H.c.} \right)
+ \frac{U}{2}  \sum_{i} \left(\sum_{a\sigma}
n_{i,a\sigma}\right)^2,
\label{hubbard}
\end{equation}
where $c_{i,a\sigma}^{\dagger}$ creates an 
electron with spin $\sigma=(\uparrow,\downarrow)$
and orbital index $a=(1,2)$  at the $ith$ site, 
and  $n_{i,a\sigma}= c_{i,a\sigma}^{\dagger}c_{i,a\sigma}$.
The total symmetry group of (\ref{hubbard}) is 
U(4) = U(1)$_{\rm {charge}}\otimes$ SU(4)$_{\rm {spin-orbital}}$, 
it  is the maximal symmetry allowed for a two-band Hubbard model.
A simple one-loop perturbative analysis \cite{lin} would predict that,  
at half-filling, a SO(8) symmetry between {\it charge} and 
spin-orbital degrees of freedom
is likely to be dynamically enlarged at low energy. 
Such a DSE pattern, U(4)  $\rightarrow$ SO(8),  
is highly non-trivial since one naturally expects the
charge degrees of freedom to decouple at sufficiently large $U$. Indeed, in the limit
$U \gg t$ the Hamiltonian (\ref{hubbard}) reduces, at half-filling, 
to an antiferromagetic (AF) Heisenberg model 
(where the  spin operators act on the 
six-dimensional antisymmetric representation of SU(4)).
It is precisely the interplay between the small-$U$ 
predicted SO(8) regime and the large-$U$ charge-decoupled
Heisenberg limit which is considered here.

The low-energy effective field theory associated 
with (\ref{hubbard}) is obtained, as usual,
by performing the continuum limit and is most 
suitably expressed in
terms of the excitations that are related to the  symmetry 
of the problem.  The U(1)$_{\rm {charge}}$ charge sector  is described, 
in a standard way,  by a single  bosonic field $\Phi_c$
and its dual field $\Theta_c$. There are many equivalent ways to describe 
the spin-orbital excitations in the SU(4)$_{\rm {spin-orbital}}$ sector
and it is  most convenient to represent them
by six real (Majorana) 
fermions  $\xi^a$, $a=(1,...,6)$ \cite{azaria}. 
We find that
the  low-energy effective Hamiltonian 
associated with (\ref{hubbard}) is given by:   
\begin{eqnarray}
{\cal H} = 
\frac{v_c}{2} \left[ \frac{1}{K_c}({\partial}_{x} \Phi_c)^2
+ K_c
({\partial}_{x} \Theta_c)^2 \right]
-\frac{iv_s}{2} \sum_{a=1}^6\left(\xi^a_R \partial_{x}
\xi^a_R \right. \nonumber \\ 
-  \left. \xi^a_L \partial_{x}\xi^a_L\right) 
+ \pi g_s\left(\sum_{a=1}^6 \kappa^a\right)^2 
- 2 i g_{sc} \cos(\sqrt{4 \pi} \Phi_c) \sum_{a=1}^6 \kappa^a,
\label{heff}
\end{eqnarray}
where  $\kappa^a =  \xi^a_R\xi^a_L$, $g_s= - g_{sc}= -U/2\pi$, and  
$v_s= v_F + g_s$,  $v_F = 2t$ being the Fermi velocity. In
Eq. (\ref{heff}), the Luttinger exponent  $K_c=1/\sqrt{1+2g_c/v_F}$
and the charge velocity $v_c=v_F \sqrt{1 +2g_c/v_F}$ depend
on  the charge coupling $g_c =3U/2\pi$. 
The low-energy effective field theory (\ref{heff}) describes
the interaction
between a SO(6)  Gross-Neveu (GN) 
model, associated with spin-orbital 
degrees of freedom, and a
Luttinger liquid Hamiltonian in the charge sector. 
The interaction term, with coupling constant $g_{sc}$, 
is an umklapp contribution that comes from the 4k$_F$  
part of the Hamiltonian density 
and is only present at 
half-filling $k_F = \pi/2$.
In sharp contrast with the half-filled SU(2) Hubbard model
and the SU(4) case at quarter-filling \cite{assaraf},  
there is {\it no} spin-charge separation at low-energy at half-filling. 
Spin-orbital and charge
degrees of freedom remain strongly coupled through 
the 4k$_F$  umklapp process.
At this point it is worth stressing that 
there exists a  higher-order umklapp term (8k$_F$ process)
${\cal V}_c =  y \,   \cos(\sqrt{16 \pi} \Phi_c)$
which depends only on the charge degrees of freedom.
Although this operator, with scaling dimension $\Delta=4K_c$, is 
strongly irrelevant at small $U$, it may  become relevant
at sufficiently large $U$. As we shall see, this contribution is 
at the heart of the physics of the SU(4) Hubbard model 
in the large $U$ limit.

A simple one-loop RG
calculation reveals that  the  couplings $g_a=(g_c,g_s,g_{sc})$ 
flow at strong
coupling. In particular, as $g_c$  blows up  at low energy $K_c$  
inevitably  decreases until
${\cal V}_{c}$ becomes relevant. Thus, the nature
of the low-energy physics depends  on the balance 
between the two umklapp operators with very different
properties.
Clearly non-perturbative 
methods are called for.
In this respect, Gerganov {\it et al.} \cite{gerganov}  
have provided  an RG framework
which allows to compute the  RG $\beta$ function {\it to all order} in
perturbation theory for a large class of one-dimensional  
models with current-current interactions.  We have applied
their formalism to the Hamiltonian  (\ref{heff}) and obtained 
the resummed  $\beta$ 
function.  
The detailed analysis of the non-perturbative RG flow will be 
discussed elsewhere \cite{longver} and we shall here present our main  result.
Neglecting  velocity anisotropy, we get:
\begin{eqnarray}
\dot{g}_c &=& 24\,(g_c-2)^2\,\frac{g_{sc}^2}{(g_{sc}^2-4)^2}
\nonumber\\
\dot{g}_s &=& 
\frac{16 g_s^2}{(g_s+2)^2}
+ 8\,(g_s-2)^2\,\frac{g_{sc}^2}{(g_{sc}^2-4)^2}
\nonumber\\
\dot{g}_{sc} &=& \frac{4g_{sc}}{4 - g_{sc}^2} \left[\,
12-(g_{sc}^2+4)\left(\frac{1}{g_c+2}+\frac{5}{g_s+2}\right)\right],
\label{rgequations}
\end{eqnarray}
where $\dot{g}_a = \partial g_a/\partial t$, $t$ being the 
RG ``time'' and $g_a \rightarrow g_a/v_F$. 
In absence of the umklapp contribution ${\cal V}_c$, we find that 
the RG flow crucially depends on $g_c$ as follows. 
In the weak-coupling regime, 
at  small enough $U/t$ such that  $g_c \leq  2$, 
all the couplings converge to the $same$ 
value, $g_a(t^*) = 2$, at some finite RG ``time''  $t^*$. 
On the other hand, when   $g_c >  2$,  one enters a regime
where perturbation theory is meaningless.

{\it Weak Coupling Regime}. When $g_c  <  2$,  
much can be said on the low-energy physical
properties of the model (\ref{hubbard}). Indeed, 
integrating  the flow up to $t^*$, one finds 
that the  Hamiltonian (\ref{heff})  at that scale reduces 
to the  SO(8) GN model: 
\begin{equation}
{\cal H}^* =  -\frac{iv}{2} \sum_{a=1}^8( \xi^a_R {\partial}_{x}\xi^a_R  
-  \xi^a_L {\partial}_{x}\xi^a_L)+
 2\pi v \, \left(\sum_{a=1}^8  \kappa^a \right)^2,
\label{SO8}
\end{equation}
where we have refermionized the charge degrees of freedom in
terms of two real fermions $\xi^{7,8}$: 
$(\xi^7 + i \xi^8)_{R(L)}
\sim \exp{(\pm i \sqrt{4\pi}\Phi_{cR(L)})}$.  
The equivalence at low energy 
between (\ref{hubbard}) and (\ref{SO8}) is a manifestation of the DSE
U(1)$_{\rm {charge}}\otimes$ SU(4)$_{\rm {spin-orbital}}$  $\rightarrow$ 
SO(8). This SO(8) enlarged symmetry which  has been first
predicted  using a 1-loop RG calculation in \cite{lin},
is shown here to hold  beyond perturbation theory provided
$g_c  <  2$.   For higher values of $g_c$, the higher-umklapp term ${\cal V}_c$ plays 
a prominent role at  low-energy and, as we shall see, is responsible of
the dynamical decoupling of the charge degree of freedom.
One of the main interest of the emergence of this SO(8) symmetry 
stems from the fact that the model (\ref{SO8}) is integrable and 
a large amount of information can be extracted from the 
exact solution \cite{lin,konik}.  
The low-lying spectrum of the SO(8) GN model (\ref{SO8}) 
is fully gapped and consists of three distinct 
octets with the same mass $m\sim t e^{-t/U}$.
The fundamental fermion octet,
associated with the Majorana fermions 
$\xi^a$ of Eq. (\ref{SO8}),
is made of 
two charged $\pm 2e$ spin-orbital singlets, called cooperons,
and six spin-orbital excitations which transform according
to the self-conjugate representation of SU(4) with dimension 6.
The remaining two octets are of kinks type.
In particular, the excitations of the SU(4) Hubbard 
model (\ref{hubbard}), carrying the 
quantum numbers of the lattice fermions $c_{i,a\sigma}$,
are represented by eight of these kinks. 
In addition, there are 28 bosonic states organized 
as a rank-2 SO(8) antisymmetric tensor and 
a singlet, all of mass $\sqrt{3} m$ which 
can be viewed as bound states of the fundamental
fermions or of the kinks states.
The massive phase corresponding to the
SO(8) GN model (\ref{SO8}) is a spin-Peierls (SP) 
phase as it can be readily shown
by considering the order parameter
${\cal O}_{SP} =   \sum_{i,a\alpha}\, (-1)^i c_{i,a\alpha}^{\dagger}
c_{i+1,a\alpha}$ which  has
a non-zero expectation value $\langle {\cal O}_{SP} \rangle \neq 0$. 
The ground state
of (\ref{SO8}) is thus doubly degenerate and 
spontaneously breaks the lattice translation symmetry. 
The striking feature is that, despite of this dimerization,  
both electronic and spin-orbital excitations
are {\it coherent}, i.e. they contribute to sharp peaks
in various spectral functions.
This result stems from the 
existence in the exact spectrum of (\ref{SO8})
of states that have the same quantum numbers as the electron
and spin-orbital operators $c^{\dagger}_{i,a\sigma}c_{i,b\beta}$.
In particular, the dynamical structure spin factor of the system
displays a sharp peak at energy $\omega = \sqrt{3} m$
corresponding to an excitation of one of the bosonic
states of the SO(8) theory. 
In this respect, the half-filled SU(4) Hubbard model is predicted
to be a fully {\it coherent} gapped dimerized liquid at weak-coupling.
\begin{figure}
\centerline{\psfig{file=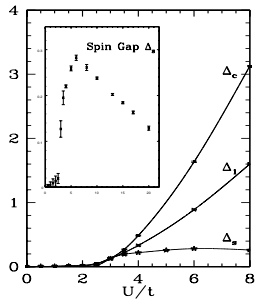,height=6cm}}
\caption{One-particle gap $\Delta_1$,
spin-orbital gap $\Delta_s$, and cooperon gap
$\Delta_c$ as a function of $U/t$;
inset: $\Delta_s$ gap as function of $U/t$.}
\label{gaps}
\end{figure} 
At this point it is worth discussing the stability of this
SO(8) phase. Although the RG
equations (\ref{rgequations}) are non-perturbative in nature 
it remains to investigate
the effect of neglected symmetry-breaking operators
such as the higher-umklapp term ${\cal V}_c$
and chiral interactions that account for velocities anisotropy. 
For small symmetry-breaking terms,  
the SO(8) multiplets will be adiabatically deformed and split into
U(1)$_{\rm {charge}}$ $\times$ SU(4)$_{\rm {spin-orbital}}$ multiplets:
the SO(8) symmetry is only realized approximately at weak 
enough coupling. 
At small $U/t$ the splittings are  exponentially small but   
we expect perturbation theory to break down as $U$ increases  
even when  $g_c  <  2$. The reason stems
from the 
neglected umklapp operator ${\cal V}_{c}$ which
becomes relevant before one reaches the SO(8) symmetry restoration point 
as  $\Delta <  2$ when  $g_c > 3/2$. We thus expect the 
SO(8) regime to hold
approximately up to some critical value $U_c$ 
of which a very naive estimate can be obtained using 
the bare value of $g_c$:
$U_c \sim 2\pi t$.

In order to check our theoretical predictions 
we have performed extensive $T=0$
QMC simulations of the SU(4) 
Hubbard model (\ref{hubbard})
at half filling for a wide range of $U/t$.
Following  the work done in  Ref. \cite{assaraf} 
in the quarter-filled case,
we have computed all gaps associated with the 
SO(8) tower of states. We discuss here
our results for three of them:
$\Delta_1$ which  is the gap to the 
one-particle excitation $c^{\dagger}_{i,a\sigma}$,
$\Delta_s$ which  is the spin-orbital  gap associated with the excitations 
$c^{\dagger}_{i,a\sigma}c_{i,b\mu}$ and finally the 
cooperon gap $\Delta_c$ which is  the gap
to a spin-orbital singlet state of  charge $2e$. The latter 
excitation is a striking
feature of the SO(8) spectrum and is not simply 
related to electronic excitations on the lattice. 
For example, the cooperon comes into  pairs  from   the   
charge $4e$ excitation   $\Pi_{a\sigma} c^{\dagger}_{i,a\sigma}$.
We have computed the cooperon gap $\Delta_c$ as half the gap 
of this state.
The exact spectrum of (\ref{SO8}) imposes the highly 
non-trivial predictions
for the ratios:  $(\Delta_1/\Delta_c)_{SO(8)} = 1$ 
and   $(\Delta_s/\Delta_c)_{SO(8)} = \sqrt{3}$.
Strong deviations from these theoretical
predictions will be a signature of the failure of the 
increased SO(8) symmetry. 
We show in Fig. (\ref{gaps}) our results for 
$\Delta_1(U)$, $\Delta_s(U)$ and $\Delta_c(U)$ for values of 
$U/t$ ranging from $0.5$  to $20$. The extrapolation to 
the thermodynamical limit has been performed
using  lattice  sizes $L=8,16,32,48,64$ and the errors on the gaps 
range from $10^{-2}$ at small $U/t$
to $10^{-3}$ at large $U/t$. Two asymptotic regimes are  
identified. A small $U/t$ regime  
and a large $U/t$ regime  where  spin-orbital and charge
degrees of freedom clearly separate. Both regimes are most easily seen on the
spin gap $\Delta_s(U)$ behavior (inset of 
Fig. \ref{gaps}) which  increases until  it reaches 
a maximum around  $U/t \sim 6$
and then decreases smoothly to zero 
as $U/t \rightarrow \infty$. Clearly the SO(8)
regime is expected to show off at small $U/t$. 
\begin{figure}
\centerline{\psfig{file=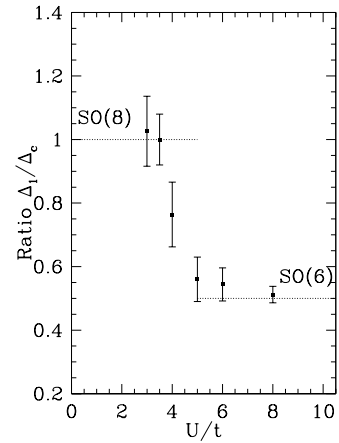,height=6cm}}
\caption{Gap ratio $\Delta_1/\Delta_c$ as a function of $U/t$.}
\label{ratio1}
\end{figure}    
In Fig.\ref{ratio1} we plot the ratio 
$(\Delta_1/\Delta_c)(U/t)$. Despite our high quality
QMC datas,  we have not been able to resolve the ratios at very small $U$'s 
where the gaps are exponentially small. However, one observes a clear  
saturation of the ratio at the SO(8) value  as $U$ decreases below 
$U \sim  3.5t$.  Other gap ratios, not presented here\cite{longver},
show  also  a SO(8)  saturation in the regime $U \lesssim 3.5t$. 
These results strongly support the existence of a  SO(8) DSE
in a regime where the gaps are not infinitesimally
small ($\Delta_s \sim 0.1t-0.2t$).
Above $U \sim 3.5 t$,  the ratio  shows 
a  departure from  its SO(8) value. Though such 
a behavior   may be attributed to   
level splitting due to symmetry breaking operators
at small $U/t$, this is certainly not the case  
above  $U \sim 6 t$ where $\Delta_1/\Delta_c$
saturates at the value $1/2$. 
It is  difficult  from our results
to give a precise value to $U_c$ above which the SO(8) regime is lost 
but  we can
give an  estimate $3 t \le U_c \le 6 t$. 
Notice that the upper value  is in agreement with our rough estimate
of $U_c = 2\pi t$ based on a  scaling argument. 
We shall see now
that the physics at large $U/t$ is of a very different nature.

{\it  Strong Coupling Regime}. 
When $U/t \gg 1$ there is a clear 
separation between spin-orbital and charge degrees since  $\Delta_c \gg \Delta_s$
(see Fig.\ref{gaps}).
The umklapp term ${\cal V}_c$, which depends only
on the charge degrees of freedom, becomes now much more relevant 
than the $4k_F$ coupling and charge fluctuations
are strongly suppressed by this process. 
Integrating out the charge degrees of freedom,
the low-energy effective Hamiltonian
in the spin-orbital sector reduces to a massive SO(6) GN model: 
\begin{eqnarray}
{\cal H}_{so} & \simeq & -\frac{iv_s}{2} \, \sum_{a=
1}^6( \xi^a_R {\partial}_{x}\xi^a_R  -  \xi^a_L {\partial}_{x}\xi^a_L)
- i M \, \sum_{a=1}^6  \kappa^a  \nonumber \\
&+& G_s(U)  \left(\sum_{a=1}^6 \kappa^a\right)^2,
\label{SO6} 
\end{eqnarray}
where 
$M> 0$ 
and $G_s(U)$ is a negative effective coupling
at large $U/t$. 
The Hamiltonian (\ref{SO6})
describes {\it six} massive  
Majorana fermions with a weak repulsion.
One can show, using Eq. (\ref{SO6}),
that $\langle {\cal O}_{SP} \rangle \neq 0$ so that the ground state is still 
in a SP phase.
Neglecting charge fluctuations, this dimerized phase, 
with broken translational symmetry, 
can be simply understood as a set of nearest-neighbor 
SU(4) $\sim$ SO(6) spin-orbital singlet bonds.
There is thus a continuity between weak and strong
coupling with respect to the nature of the ground state.
However, there is a striking difference between
the SO(8) regime and this strong coupling phase,
called SO(6) regime,
at the level of the coherence of excitations.
The excitation spectrum 
of the model (\ref{SO6}) for $G_s <  0$ consists 
of massive fermions $\xi^a$,
which are the SU(4) dimerization kinks,
and their multiparticle excitations. 
In particular, there are no bound states 
so that the spin-orbital 
dynamical structure factor
exhibits a two-particle continuum: 
the spin-orbital excitations are now {\it incoherent}. 
Apart from these neutral excitations, 
there are massive modes corresponding to
solitons in $\Phi_c$ with charge $q=\pm e$
coupled with zero modes of the Majorana fermions $\xi^a$
of Eq. (\ref{SO6}).
These excitations have a larger gap and carry the same
quantum numbers as the 
kinks of the SO(8) spectrum \cite{longver}.
The cooperon is no longer a stable excitation
in the large $U/t$ limit but becomes instead a diffusive
state made of these two kinks.
One thus expects that the gap ratio $\Delta_1/\Delta_c$
saturates at $1/2$ in the SO(6) regime in full
agreement with the numerical results of Fig. (\ref{ratio1}).
The physics of the strong-coupling regime can also
be investigated by a complementary approach which 
consists to map directly, in the large $U$ limit,
the SU(4) Hubbard model (\ref{hubbard})
onto a SU(4) AF Heisenberg 
chain by a standard perturbation theory in $t/U$ \cite{onufriev}:
${\cal H}_{\rm eff} = J \sum_i {\cal S}_i 
\cdot {\cal S}_{i+1}$,  
with $J = 4t^2/U$
and ${\cal S}^A_i$ are SU(4) spin which
belongs to the six-dimensional representation of 
SU(4).
This SO(6) AF Heisenberg chain
is not integrable and
has been studied by means of the density matrix
RG approach \cite{onufriev}. 
In full agreement with our results,
this model
belongs to a SU(4) dimerized phase.
Using the numerical results of Ref. \cite{onufriev},
we find that our QMC results for the spin gap 
$\Delta_s(U/t)$ follow
a SO(6) Heisenberg regime for $U > 8t$.
In this respect,
we deduce that
the low-energy physics of the SO(6) AF Heisenberg chain
is described by the
six almost free massive Majorana fermions (\ref{SO6}).

{\it  Cross-Over Regime}. Both SO(8) and SO(6) regimes differ by 
the coherent nature of spin-orbital excitations
and the existence of an elementary charge $2e$ cooperon excitation.
In the simplest hypothesis, the cross-over between these two regimes can 
be understood as a change of sign of the coupling $G_s$ as a function of $U$.
A mean-field analysis of the low-energy effective theory together
with our numerical result predicts that such a cross-over occurs
at  $U \simeq 4.5 t$ \cite{longver}.
When $G_s(U) > 0$, the Majorana fermions
of Eq. (\ref{SO6}) experience an attractive interaction
and neutral bound-state in the adjoint
representation of SU(4) are formed. 
The latter excitation is adiabatically connected 
to one of the bosonic states of the SO(8) spectrum which
is responsible of the sharp peak in the dynamical structure factor
in the SO(8) regime.  It is thus  very tempting to conclude, within
this simple scenario, that the SO(8) regime approximately extends up to $U_c\sim 4.5 t$
above which one enters the Heisenberg  SO(6) regime.

\end{document}